\documentclass[twocolumn,pre,aps,showpacs]{revtex4}
\usepackage{graphicx}

\def\be{\begin{equation}}
\def\ee{\end{equation}}
\def\bea{\begin{eqnarray}}
\def\eea{\end{eqnarray}}

\begin{document}
\title{Temperature Relaxation in Hot Dense Hydrogen}
\author{Michael S. Murillo$^1$ and M. W. C. Dharma-wardana$^2$}
%\email{murillo@lanl.gov}
\affiliation{$^1$Physics Division, Los Alamos National Laboratory, Los Alamos, New Mexico 87545\\
$^2$National Research Council, Ottawa, Canada K1A 0R6}
\email{chandre@argos.phy.nrc.ca}
\date{\today}

\begin{abstract}
Temperature equilibration of hydrogen is studied for conditions relevant to inertial confinement
fusion.  New molecular-dynamics simulations and results from quantum many-body theory are compared with
Landau-Spitzer (LS) predictions for temperatures $T$ from 50 eV to 5000 eV, and densities with Wigner-Seitz radii
$r_s = 1.0$ and $0.5$.  The relaxation is slower than
 the LS result, even for temperatures in the keV range, but converges to  agreement in the 
high-$T$ limit. 
\end{abstract}

\pacs{52.25.Kn,71.10.-w,52.27.Gr}
% 52.25.Kn  Thermodynamics of plasmas
% 52.25.Gj  Fluctuation and chaos phenomena
% 71.10.-w  Theories and models of many-electron systems
% 52.27.Gr  Strongly coupled plasmas

\maketitle

%%%%%%%%%%%%%%%%%%%%%%%%%%%%%%%%%%%%%%%%%%%%%%%%%%%%%%%%%%%%%%%%%%%%%%%%%%%%%%%%
{\it Introduction --}
While a first-principles description of the equilibrium properties of strongly coupled Coulomb systems
is a formidable task \cite{EOS}, nonequilibrium systems pose an even greater challenge. Short-pulse 
lasers and shock waves create nonequilibrium states. Thus, Coulomb systems as diverse as warm dense 
matter \cite{Riley}, ultracold plasmas \cite{Killian}, shocked semiconductors \cite{Ng}, and dense 
deuterium \cite{Knudsen} can now be readily created in the laboratory, but initially under
nonequilibrium conditions.  Similarly, energy relaxation (ER) in astrophysical plasmas is important
 to the physics of fusion of H, C, etc., in determining stellar evolution\cite{deWitt}.
 Here, we consider the ER of nonequilibrium dense hydrogen due to 
its importance in inertial confinement fusion (ICF)\cite{ICF}. 

The earliest theories of  ER in  plasmas were formulated by Landau 
\cite{Land} and Spitzer \cite{Spit} (denoted LS).  The LS approach is applicable to 
dilute, hot, fully  ionized plasms where the collisions are weak, binary, and involve negligible 
quantum effects; essentially, LS is Rutherford's Coulomb scattering formula applied to a Maxwellian
distribution. A characteristic feature of the LS approach is the 
use of a Coulomb logarithm (CL), i.e.,
\be
 {\cal L} \equiv \ln\Lambda \sim \int_{b_{min}}^{b_{max}}\!db/b \sim
\int_{k_{max}}^{k_{min}}\!dk/k
\ee
where $b_{min}$ (or $1/k_{max}$) and $b_{max}$ (or $1/k_{min}$) are suitable,
 but {\it ad hoc}, impact parameter (or momentum) cutoffs for the Coulomb 
collision. The full quantum mechanical method, based on calculating a transition rate,
does not suffer from this problem. Calculations at the Fermi golden-rule
 level and beyond have been made by Dharma-wardana et al. \cite{DWP,mwcd, Hazak}.
 Such methods automatically include degeneracy effects, effects of collective modes, and strong coupling.
Other approaches employ convergent kinetic equations
\cite{Gould_Dewitt,GMS}.  Hansen and McDonald \cite{HM} (denoted HM) and Reimann and Toepffer have directly
 obtained ER via molecular 
dynamics (MD) simulations.

In a hydrogen plasma the particle charges $z_i,z_j$ are $\pm1$, 
in atomic units, where the electronic charge $|e|$=$\hbar$=$m_e$=1. 
The mean electron and proton densities $n$ and  $\rho$ are identical. 
The ratio of a typical Coulomb energy to the kinetic energy becomes, in the classical 
regime, $\Gamma=1/(r_sT)$, where $T$ is the temperature in energy units, and $\Gamma=r_s$
in the quantum regime. 
$r_s=\left[3/(4\pi n)\right]^{1/3}$ is the radius of the Wigner-Seitz sphere of an
electron or a proton.
The properties of partially degenerate fully-ionized plasmas
require two independent parameters, e.g., both $r_s$ and $\theta=T/E_F$, 
where $E_F=\left(3\pi^2 n\right)^{2/3}/2$ is the
Fermi energy \cite{Murillo_tut}.
 The LS analysis of ER is in terms of e-p
collisions in a Maxwellian gas.
Relative to the LS approach, some theoretical approaches relax faster \cite{GMS},
or slower \cite{DWP,Hazak}. HM concluded that many-body effects are negligible
and that the the LS result holds\cite{HM}.
There is currently no 
direct experimental data for ER rates, although such experiments 
are under way\cite{Taccetti,GaretaRiley}.

Here, we present larger HM-like molecular-dynamics (MD) simulations and  
 quantum many-body calculations to narrow the gap in our predictions of ER rates for hot, dense hydrogen 
relevant to ICF targets composed of dense cryogenic fuel rapidly laser compressed to keV temperatures.
The modeling of these dense plasmas covers physical processes over many orders of magnitude in density 
and temperature. Nonequilibrium quantum simulations require theoretical
breakthroughs that are not yet fully established; however, the MD techniques
that employs quantum-corrected effective potentials can be usefully applied to these problems.
Since MD simulations attempt to solve the many-body equations of motion exactly, they are 
likely to provide accurate ER rates for hot, dense hydrogen.  

{\it Molecular dynamics.--}
Several issues arise in simulations of plasmas with temperatures in the 
$10^2$-10$^3$ eV range.  Because the screening length and mean-free-path
 are larger at higher temperatures, 
we have varied the number of particles widely (maximum of several thousand), with $N$=500 used for
the results presented here.  Also, because the ER time varies roughly as $T_e^{3/2}$, these simulations 
required millions of time steps for the higher-temperature cases.  Compounding the longer runs was 
the need for much smaller time steps (as small as $0.005\omega_{pe}^{-1}$, where $\omega_{pe}$ is the 
usual electron plasma frequency), as a result of high velocity collisions at high temperature. Such 
issues are not unexpected, but have rarely been dealt with, since MD is typically applied to strongly 
coupled (i.e., cooler) classical  systems.  We carried out direct simulations of electrons 
and protons since the Born-Oppenheimer approximation is not applicable to this problem. The  
mass ratio ($1:1836$) was used due to our interest in mass-dependent collective mode effects.
Finally, integrations were carried out with the velocity-Verlet algorithm using an ${\cal O}(N^{3/2})$ 
Ewald method.

The Coulomb interaction $1/r$ of classical physics is replaced by the mean-value of the
operator $1/\hat{r}$ in quantum systems. This feature modifies the short-ranged
behaviour of the electron-electron and electron-proton interactions, since the
 de Broglie wavelength
of the electron is not negligible for small $r$.
 Also, unlike the classical electron-proton
interaction which always leads to a bound state, the delocalized electron does not bind to
the proton for regimes studied here. These effects are included in the
e-e and e-p potentials by solving the relevant Schrodinger equations for the two-body scattering
 processes. These lead to diffraction-corrected Coulomb potentials 
$v_{ij}^{dfr}(r_{ij})$
 which are Coulomb-like for
distances larger than the respective de Broglie lengths,
 $\lambda_{ij}=1/\sqrt{2\pi\mu_{ij}T_e}$.
Here $\mu_{ij}$ are the effective masses of the colliding pair. 
The choice of $b_{min}=\lambda_{ij}$  is usal in
the LS approaches.

Then interaction $\phi_{ij}(r)$ between the pair $i,j$ is given by
\bea
\label{phys_mod_dfr}
\phi_{ij}&=&v_{ij}^{dfr}(r_{ij})+v_{ee}^{Pau}(r)\\
v_{ij}^{dfr}(r_{ij})& =& \frac{z_iz_j}{r_{ij}}\left[1-e^{\left(-\frac{f_{ij}r_{ij}}{
\lambda_{ij}} \right)} \right] + v^{Ewl}_{ij}(r_{ij})
\eea
where $v^{Ewl}_{ij}$ is the Ewald potential. 
The potential, $v_{ee}^{Pau}$,
accounts for the spin-averaged Pauli exclusion  between two electrons. 
This ensures that the ``non-interacting'' electron pair distribution function (PDF) 
calculated from a classical simulation
  is just the non-interacting quantum PDF \cite{Lado, CHNC}.
The explicit form\cite{JonesMurillo} used by HM is adequate
 for most of the range studied
in this paper.
%\be
%\label{phys_mod_Pauli}
%v_{ee}^{Pau}(r_{ee}) =
% -T_e\ln\left[1-\frac{1}{2}\exp\left(-\frac{r_{ee}^2}{\pi\lambda^2_{ee}}
% \right) \right].
%\ee
%This potential of Eq.~\ref{phys_mod_dfr}, (Model I), with $f_{ij}=1$, 
%is the diffraction-corrected potential 
%\cite{JonesMurillo} (used by HM). Model II is
%the complete Pauli potential \cite{Lado}.  
%Model II is essential for small $\theta<1$,
%and is the same as the Pauli potential implemented
%in the classical-map hypernetted-chain (CHNC) 
%approach \cite{CHNC}. 
The factor $f_{ij}$ is unity for all except the
cross-species case $f_{ep}$. This is chosen  using CHNC.
The CHNC uses the above potentials and a classical fluid
temperature $T_{cf}=\surd{T_e^2+T_q^2}$
 where the quantum temperature $T_q$ is defined in Ref.~\cite{CHNC}.
The $f_{ep}$ factor at a given $r_s,\, T_e,\, T_p$ 
is fixed by requiring that the  CHNC  $g_{ep}(r)$, at $r=0$
is the same as the Kohn-Sham value of  $g_{ep}(r=0)$, as discussed
more fully in Ref.~\cite{DW-Murillo}.
The factors $f_{ep}$ are within 5\% of unity for the 
conditions of our study.

\begin{figure}[t]
\includegraphics[angle=0, width=3in]{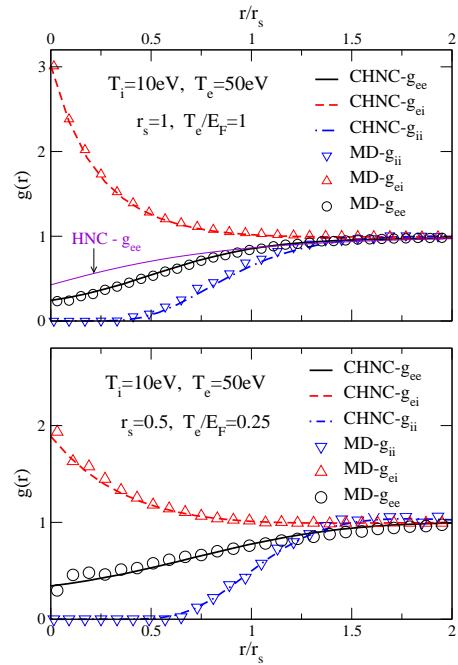}
\caption{Pair distribution functions from MD (data points), CHNC (thick lines)
for dense hydrogen, $r_s=0.5$ and 1, $T/E_F=$0.25 and 1. The $g_{ee}$
from a simple HNC calculation (i.e., no quantum temperature, no Pauli potential)
with just $v_{ij}^{dfr}$ is also shown for $r_s=1$.}
\label{rdf_fig}
\end{figure}

To validate the potentials under two-temperature, dense-plasma conditions, we 
carried out  (using $\phi_{ij}$ and its simplified HM forms)
 MD and CHNC for the conditions $r_s=1.0$, 
$T_e=50$eV, and $T_i=10$eV.  In the MD, electron and proton velocity-scaling thermostats
 were used to 
create the two-temperature system, whereas  recent results for the cross-species 
 temperature $T_{ep}$
\cite{DW-Murillo} were used in the CHNC.  The pair distribution
 functions $g_{ij}(r)$ are shown in 
Fig. (\ref{rdf_fig}).

Temperature equilibration rates were determined for the two densities $r_s=0.5$ 
($n_e=1.5\times 10^{25}$cm$^{-3}$) and $r_s=1.0$ ($n_e=1.61\times 10^{24}$cm$^{-3}$) over 
the temperature range $0.25<\theta<20$.  In practice, an equilibration 
stage with separate electron and proton velocity-scaling thermostats
 was used to establish a
two-temperature system.  This was followed by a microcanonical evolution
 in which the temperatures
were relaxed, with
\be
T_j(t)=\frac{m_j}{3N_j}\sum_{i=1}^{N_j} v_j^2(t),
\ee
for each species $j$. Energy conservation was carefully monitored to assure stability
 at the elevated temperatures.
Stability was quantified by $\Delta E = \frac{1}{N}\sum_{i=1}^N\left|\frac{E_i-E_0}{E_0}\right|$,
where $E_i$ is the energy at the $i$-th timestep.  The timestep was chosen by first performing 
several simulations with varying timesteps for temperatures $T=100, 250, 500, 750$eV and 
noting the impact on $\Delta E$. As mentioned above, the timestep required for stability 
decreases dramatically with increasing temperature.  A fit to the slope of the temperature profiles 
yielded the equilibration rate; in practice, the ion temperature was fixed for all runs at
$T_i=10$eV so that the reported results are $\partial T_e/\partial t$.  Over the range of 
temperatures considered, the relaxation rate should be very insensitive to the ion temperature.

{\it Quantum transition rates.--}
The most transparent, strictly quantum approach to the calculation of the
 ER rate is to treat it as a 
transition rate where an electron in an initial momentum state
 $\vec{k}_i$ transfers to a final state
$\vec{ k}_f$, while a proton in the initial momentum 
eigenstate $\vec{p}_i$ absorbs energy and transfers to a final
eigenstate $\vec{p}_f$. The availability of such states depends on the products of Fermi 
occupation factors $n_{k_i}(1-n_{k_f})$,
and similarly for the proton states. The strength of the transition depends
 on the matrix element between
the initial and final states. This matrix element may be taken in
 lowest-order theory (Born approximation)
or in higher order (i.e, a $T$-matrix evaluation).
 These are the usual ingredients of the Fermi golden rule (FGR) for the
transition rate. The summation over all such pair processes
 gives the total ER rate. But such summations
immediately convert the description of the plasma into a
 description containing the full
 spectrum of single-particle and collective modes. The spectrum of 
all modes is given by the spectral
function $A_j(q,\omega,T_j)$ where the species index $j=e$ or $p$.
These spectral functions are given by the imaginary parts of the
corresponding dynamic response functions $\chi^{j}(\vec{k},\omega)$, e.g,
Eq.~(16) of Ref.~\cite{Hazak}.
 The ER rate evaluated within the Fermi golden rule, $R_{fgr}$
 can be expressed in terms of 
the response functions of the plasma as follows, given in Eqs. (4)-(7)
 of Ref.\cite{mwcd}, and Eq.~(15) of Ref.~\cite{Hazak}:
%\bea
%A_j(q,\omega,T_j)&=&-2\Im \chi_{jj}(q,\omega,T_j)\\
%\chi_{jj}(q,\omega,T_j)&=&\chi_{jj}^0/\{1-V_{jj}(q)[1-G_{jj}(q,\omega)\chi_{jj}^0]\}\\
%R_{fgr}&=&\Sigma_{q,\omega}|V_{ep}|^2\omega A^e(q,\omega)
%A^p(q,\omega)\Delta N_{ep}(\omega)\\
%\Delta N_{ep}(\omega)&=&N(\omega/T_e)-N(\omega/T_p)\\
%N(\omega/T_j)&=&1/[exp^{\omega/T_j}-1]
%\eea
\bea
\label{cothform}
R_{fgr}&=&\frac{dK}{dt}=\int \frac{d^{3}k}{\left( 2\pi \right) ^{3}} 
\frac{\omega d\omega}
{2\pi}(\Delta B) F_{ep}\\
\Delta B&=&\coth(\omega/2T_e)-\coth(\omega/2T_p)\\
%&\left[ \coth \left( \frac{ \omega }{2T_{e}}\right) -
%\coth
%\left( \frac{ \omega }{2T_{p}}\right) \right]\\ 
F_{ep}&= &|( V_{ep}(k)
| ^{2}\Im\left[ \chi ^{p}(\vec{k},\omega )\right] \Im
\left[ \chi ^{e}\left( \vec{k},\omega \right) \right]
\eea
In the above we have used the spherical symmetry of the plasma
 to write scalars $q,k$ instead of
$\vec{q},\, \vec{k} $ to simplify the notation.
\begin{figure}[t]
\includegraphics[angle=0, width=3.3in]{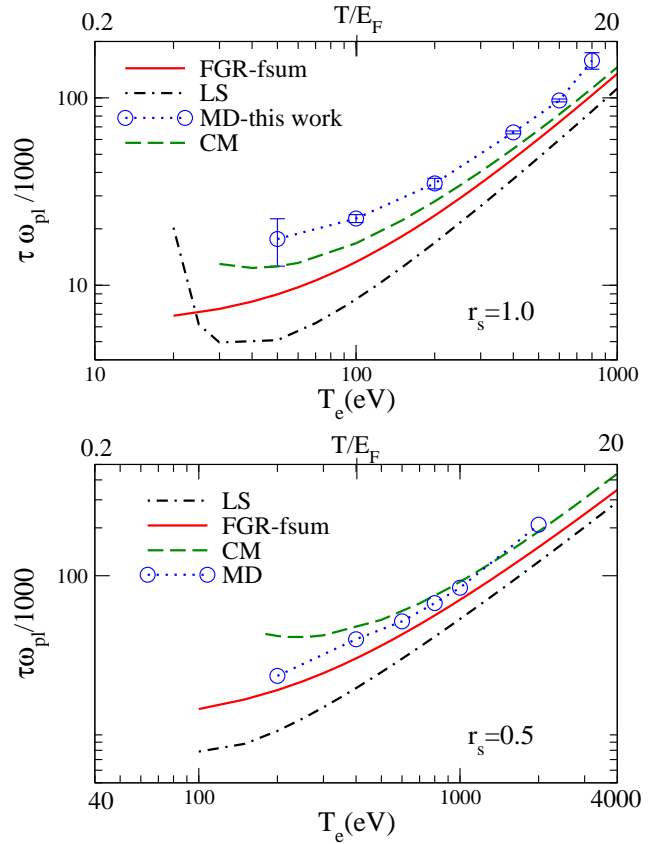}
\caption{The relaxation time $\tau/1000$,
in units of the inverse electon-plasma frequency,
for dense hydrogen, $r_s=0.5$ and 1, from the
degenerate to the classicle region. The Landau-Spitzer (LS) result
using the Hansen-McDonald(HM) prescription for the $k$-cutoffs,
the results from the Fermi Golden rule(FGR), and the coupled-mode(CM)
are shown, together with the MD results.
}
\label{taugraph}
\end{figure}

The excess-energy density is denoted by $K=K_e-K_i$, and becomes $3(T_e-T_p)n/2$,
in the classical regime, i.e., where the chemical potential $\mu$ is negative.
The interaction $V_{ep}(k)$ in this equation is the full Coulomb matrix element
and {\em not} the diffraction corrected form used in the CHNC and the
classical simulations. In the simplest form of the FGR, $V_{ep}(k)=4\pi/k^2$ since
the momentum states are taken to be plane waves. A $T$-matrix evaluation would use
phase-shifted plane waves and the corresponding modified density of states, instead
of $d^{3}k/\left( 2\pi \right) ^{3}$.
With the onset of the classical regime where
$\mu<0$, which occurs for $\theta>1$, the $\Delta B$ factor 
 becomes $2\Delta/\omega$ where $\Delta =(T_e-T_p)$. It was shown in
Hazak et al.\cite{Hazak} how we may do the $\omega$-integration by
exploiting the $f$-sum rule and the fact that the ion-spectral function,
peaking near the ion-plasma frequency, resides way below the electron
spectral function. Then Eq.~\ref{cothform} can be written, to a good approximation
as:
\bea
\frac{1}{\Delta}\frac{d\Delta}{dt}&=&\frac{2}{3n}\omega _{ion}^{2}\int_0^\infty\frac{2}{\pi} 
\left[ \frac{\partial }{\partial \omega }\Im\chi ^{ee}\left(
k,\omega \right) \right] _{\omega =0}dk
\label{fgrBA}
\eea
where $\omega _{ion}$ is the proton-plasma frequency. If we keep the proton
 temperature $T_p$ fixed, we see that Eq.~\ref{fgrBA} leads to a
relaxation time $\tau$ for the electron temperature
$T_e$, involving the inverse of the r.h.s. of Eq.~\ref{fgrBA}.

The above analysis treats the plasma as two independent subsystems. In reality,
the ion-density fluctuations are screened by the electron subsystem, and the
ion-plasma mode becomes an ion-acoustic mode. The excitations in the
coupled-mode system are described by the zeros of Eq.~(45) of Ref.~\cite{DWP}.
In the static, $k\to0$ limit this denominator converts the electron screening
parameter $k^e_{DH}$ to $\surd\{(k^e_{DH})^2+(k^p_{DH})^2\}$. However, the
proton-density modes act dynamically in the relaxation process.
 Thus the use of static ion screening
is incorrect. The  coupled-mode(CM) approach is fully dynamical and
includes another denominator,
\be
\label{dcm}
d_{cm}=|1-V_{ep}^2(k)\chi ^{ee}\left(k,\omega \right)\chi ^{pp}\left(k,\omega \right)|^2
\ee
into the integrand in Eq.~\ref{cothform}. That is, $F_{ep}$ in Eq.~\ref{cothform} is
replaced by $F_{ep}/d_{cm}$.
 
The simple Landau-Spitzer form can also be be written in the same form as Eq.~\ref{fgrBA},
as shown in Ref.~\cite{Hazak}.
 The quantum approaches in CM and FGR automatically contain the diffraction and screening
effects. Thus, while Eq.~\ref{fgrBA} use the full integration $0\to\infty$,
LS needs the cutoffs $k_{min}$ and $k_{max}$ to obtain a convergent result.
The calculated LS-values of $\tau$ does depend somewhat
on the choice of $k_{min}$ and $k_{max}$. Hence 
 different realizations
of the LS-form need not reduce to the same result  at finite $T/E_F$.
In fact  Lee and More\cite{leemore} use cutoffs based on the full
static screening length which includes the ions as well and
differ significantly from LS.
 However, the CM analysis clearly
shows that the ion response in ER is dynamic.

The  non-interacting response function $\chi^0(q,\omega,T)$ at arbitrary
degeneracies was given by Khanna and Glyde\cite{KH}. We use a generalized RPA form
where local filed corrections $G_{ee}(k)$ may be included\cite{DWP}. However, these
are quite small for the conditions of this study.
Both the FGR and CM calculations assume that linear response can be used to
discuss the interaction of a proton with the electrons. The resulting
calculations are shown in Fig.~\ref{taugraph}.
The coupled-mode (CM) calculation is
quite close to the Fermi golden rule (FGR)
f-sum result. This is expected since
the H-plasmas considered here are relatively
 weakly coupled, with $\Gamma<1$. Nevertheless, the inclusion of CM leads to
 better agreement with the MD simulation. 
Also, we have assumed the bare $4\pi/k^2$ form for the $V_{ep}$ in the
FGR and CM formulae,
without the moderating effects of a pseudo-potential. Such effects
would tend to make the $\tau$ larger than that from the present FGR or CM calculation.
This linear-response assumption is more satisfactory for the $r_s=0.5$ plasma.
Thus the MD results at $r_s=0.5$ are very close to the CM results.

{\it Conclusion.--} We have evaluated the temperature relaxation time in hot,
 dense hydrogen using 
state-of-the-art molecular dynamics simulations and quantum
many-body theory.  We find that the relaxation is slower than the LS value even for
temperatures in the kilovolt range, which suggests that burning plasmas are slightly more out
of equilibrium that might have been expected.  Unlike in the calculations presented in, say, 
Ref.~\cite{DWP,mwcd}, where strongly coupled Al-plasmas were considered, the present calculations
 are for systems with $\Gamma \sim 1$ or less. 
 The temperatures have been pushed to $T/E_F\simeq20$. Thus we see
that the CM, FGR and the LS forms converge for sufficiently large $T/E_F$.
The MD results are slightly higher than from the analytical models which
use linear response.  It is also clear that 
the LS form is inadequate  for highly compressed low-$T$,
 partially degenerate plasmas. 

%%%%%%%%%%%%%%%%%%%%%%%%%%%%%%%%%%%%%%%%%%%%%%%%%%%%%%%%%%%%%%%%%%%%%%%%%%%%%%%%

%%%%%%%%%%%%%%%%%%%%%%%%%%%%%%%%%%%%%%%%%%%%%%%%%%%%%%%%%%%%%%%%%%%%%%%%%%%%%%%%
%%%                              FIGURE CAPTIONS                             %%%
%%%%%%%%%%%%%%%%%%%%%%%%%%%%%%%%%%%%%%%%%%%%%%%%%%%%%%%%%%%%%%%%%%%%%%%%%%%%%%%%

\end{document}